%% file: main.tex
\documentclass[conference]{IEEEtran}

\usepackage{booktabs} 
\usepackage{graphicx}
\usepackage{float}
\usepackage{dblfloatfix}
\usepackage{subcaption}
\usepackage{bm}
\usepackage{multirow}
\usepackage{threeparttable} 
\usepackage{hyperref}
\usepackage{fancyvrb}
\usepackage[usenames, dvipsnames]{color}

\usepackage{color}
\usepackage{soul}

\definecolor{bggreen}{RGB}{186,201,186}
\definecolor{codegreen}{rgb}{0,0.6,0}
\definecolor{codegray}{rgb}{0.5,0.5,0.5}
\definecolor{codepurple}{rgb}{0.58,0,0.82}
\definecolor{backcolour}{rgb}{0.95,0.95,0.92}
\definecolor{lightgray}{gray}{0.90}

\begin{document}
\title{Adapting Neural Text Classification for\\Improved Software Categorization}

\author{
\IEEEauthorblockN{Alexander LeClair, Zachary Eberhart, and Collin McMillan}
\IEEEauthorblockA{University of Notre Dame, 
	              Department of Computer Science, 
	              Notre Dame, IN 46656\\
	              Email: \{aleclair, zeberhar, cmc\}@nd.edu}
}

\maketitle

\begin{abstract}
Software Categorization is the task of organizing software into groups that broadly describe the behavior of the software, such as ``editors'' or ``science.''  Categorization plays an important role in several maintenance tasks, such as repository navigation and feature elicitation.  Current approaches attempt to cast the problem as text classification, to make use of the rich body of literature from the NLP domain.  However, as we will show in this paper, text classification algorithms are generally not applicable off-the-shelf to source code; we found that they work well when high-level project descriptions are available, but suffer very large performance penalties when classifying source code and comments only.  We propose a set of adaptations to a state-of-the-art neural classification algorithm, and perform two evaluations: one with reference data from Debian end-user programs, and one with a set of C/C++ libraries that we hired professional programmers to annotate.  We show that our proposed approach achieves performance exceeding that of previous software classification techniques as well as a state-of-the-art neural text classification technique.
\end{abstract}

\IEEEpeerreviewmaketitle

\input{intro}
\input{problem}

\input{background}

\input{dataprep}

\input{approach}

\input{eval}

\input{evalresults}
\input{discussion}

\vspace{-0.1cm}
\section{Reproducibility}
\label{sec:reproducibility}
To encourage further research and to facilitate replication of our results, we provide all scripts, raw and processed data, experimental input and output, and other information via an online appendix.  As with any paper relying on machine learning algorithms, the number of parameters, software versions, and other model configurations far exceeds what can be explained in a single manuscript.  We have tried to condense the results to only those necessary to the objective of this paper, but we include several other supporting configurations:
\texttt{\url{http://173.255.251.249/codecat/}}

\section*{Acknowledgment}
This work is supported in part by the NSF CCF-1452959, CCF-1717607, and CNS-1510329 grants. Any opinions, findings, and conclusions expressed herein are the authors’ and do not necessarily reflect those of the sponsors.

\vfill

\bibliographystyle{IEEEtran}
\bibliography{main} 

\end{document}

%% file: intro.tex
\section{Introduction}
\label{sec:intro}

Software Categorization is the task of organizing software into groups that broadly describe the behavior of the software, for example, \texttt{sound} or \texttt{web}~\cite{linares2014using}.  Categorization has long been understood to play an important role in software maintenance, for tasks such as helping programmers locate programs in large repositories of code~\cite{Sharma:2017:CGR:3084226.3084287, McMillan:2011:CSA:2117694.2119646, Ugurel:2002:WCA:775047.775141, wang2013mining}, identifying features to prioritize~\cite{frakes1998dare, kang2002feature}, and finding similar programs to one's own.  For example, McMillan~\emph{et al.}~\cite{McMillan:2012:DSS:2337223.2337267} describe a situation in which stakeholders must find similar programs during feature elicitation in a competitive commercial bidding process.

The problem of software categorization is usually defined as a coarse-grained categorization of whole projects.  Sometimes this process is supervised, such as the assignment of a set of known tags to code repositories~\cite{Sharma:2017:CGR:3084226.3084287, thung2012detecting} or categories to a dataset of projects~\cite{Ugurel:2002:WCA:775047.775141, linares2014using}.  Unsupervised procedures are popular as well, with LDA being a particularly important tool for grouping similar programs~\cite{wang2012labeled, wu2016tag2word, tian2009using, baldi2008theory, kawaguchi2006mudablue, vargas2015automated}.  In a rich variety of ways, the usefulness of automatic classification and calculation of similarity of programs is well established in the literature.

A logical choice in attempting automatic software categorization is to turn to the existing literature in text classification.  Text classification is a well-studied research area, and quite excellent results are reported for tasks such as sentiment analysis and news article classification (see Section~\ref{sec:background}).  But as we will show in this paper, these approaches are generally not effective off-the-shelf for software classification.  The reason, in a nutshell, is the vocabulary problem in source code that has been observed for decades~\cite{Biggerstaff:1993:CAP:257572.257679, Marcus:icse2003, Hill:2009:ACS:1555001.1555039}: the terms used at a low-level in source code tend to vary more than terms in text documents due to programmer-generated identifier names, the widespread use of abbreviations and compound words that may be hard to expand or split, and the specific meanings that many words have in code that are different than their English meanings (e.g. button, free, object).  In short, the low-level code details do not align with high-level concepts~\cite{Biggerstaff:1993:CAP:257572.257679}.

There are signs prior to this paper that the vocabulary problem in software affects automatic categorization.  A revealing result by Wang~\emph{et. al}~\cite{wang2013mining} at ICSM'13 showed quite good performance in placing software into hierarchical categories.  The result is informative because they achieved it by mining the software's profile including online resources and other text data -- they did not rely solely on the source code.  In contrast, Linares~\emph{et. al}~\cite{linares2014using} reported at ICSM'11 and EMSE far lower performance when using attributes from source code only.  Notably, both sets of authors used the bag-of-words model and a linear classifier, and a dataset derived from Java projects downloaded from SourceForge.  (However, there are a few important caveats we discuss in Section~\ref{sec:background}.)

Unfortunately, it is often not realistic to assume that software has an online profile, a long text description, or significant high-level documentation.  New projects added to an open-source repository may have only the source code available (a problem GitHub has faced during automatic tagging~\cite{github:categoryblog}), and commercial repositories may have many projects with only limited related text information such as legacy code~\cite{McMillan:2012:DSS:2337223.2337267}.  We cannot assume that we will have high-level text data available in these situations; if we wish to categorize a project that does not have this information, we will have to rely on the knowledge encoded into the source.

In this paper, we present an approach for using pre-trained word embeddings in a neural architecture with several adaptations for software categorization.  Essentially what we propose is a procedure for pre-training word embeddings that we designed to encode the similarity between low-level terms in the source code to high-level terms in project descriptions. These high-level descriptions are sometimes available in large repositories of projects.  Then we built a neural classification architecture that uses the word embedding, and trained this architecture on a large corpus of software projects (\emph{with} descriptions).  We then use the model to classify a test set of projects that have source only (\emph{no} text descriptions).

Our approach achieves results superior to a bag-of-words/linear regression (BoW+LR) baseline used by related work~\cite{wang2013mining, linares2014using} when text descriptions are available, and exhibits far better performance when high-level text information is not provided.  We found that this improvement is due to our proposed procedure for constructing word embeddings from descriptions and code on the training set: performance was limited when we tested an embedding trained on non-software data (e.g. Wikipedia) as is common in the NLP research area.  The embedding we recommend connects words in project descriptions to terms in the source code.

To give a general picture of the problem, consider that the baseline BoW+LR used by previous work achieved 68\% precision 64\% recall in a six-category classification problem in this paper, when text descriptions were available -- results similar to those reported at ICSM'13~\cite{wang2013mining}.  But when only code is available, baseline performance drops to 43\% precision 33\% recall.  The baseline is not applicable to code off-the-shelf.

We observe the same problem using a state-of-the-art text classifier based on recurrent and convolutional neural networks: using a word embedding pre-trained on a very large English corpus (provided by Pennington~\emph{et. al}~\cite{pennington2014glove}) that has been shown to perform well on natural language, we found 80\% precision 77\% recall when project descriptions were available.  But, only 53\% precision 46\% recall when only code was available.  There is an improvement over the BoW+LR baseline on the text data, as we expect given the neural technique's birthplace in NLP literature.  But we observe only marginally improved performance on the same dataset of projects, when we use only the source code.

The approach we propose in this paper achieves 86\% precision 80\% recall when text descriptions are available, which drops to 61\% precision 52\% recall when using only source code.  This is still a substantial drop, and we do not claim to ``solve'' the vocabulary problem.  However, we view the across-the-board improvements, including 6-18\% increases in precision and recall, as significant progress towards adapting text NLP techniques to a software maintenance problem.

To promote reproducibility and further study, we make our entire dataset, implementation, and evaluation materials and results available via an online appendix (see Section~\ref{sec:reproducibility}).

%% file: problem.tex
\section{Problem and Overview}
\label{sec:problem}

We target the problem of software categorization.  In keeping with the traditional problem definition, a ``project'' may be one program or a suite of programs intended to perform a particular task, and we classify these into one of several categories defined by the software repository.  To limit the scope of this one paper, we explore only single-class classification; we do not consider multi-class or hierarchical categorization at this time.  Also, we focus on categories defined by a repository, instead of unsupervised label generation.

A solution to this problem would have several applications in the short run, as discussed in the Introduction.  But in the long run, we are positioning this paper to advance the state-of-the-art in automatic program comprehension generally: much effort in the software maintenance subarea has been dedicated to automated understanding of code changes, artifacts, etc., with the hope of ``teaching the computer'' to recognize high-level rationale similar to what a human might, rather than low-level details only.  In addition, this paper contributes to an ongoing debate in Software Engineering research as to whether neural architectures are an appropriate tool given the unique constraints present in SE data~\cite{hellendoorn2017deep, allamanis2017survey}.  We argue that they are for the problem of categorization, especially given their ability to model embeddings.

\begin{figure}[!h]
	\centering
	\vspace{-0.2cm}
	\includegraphics[width=0.35\textwidth]{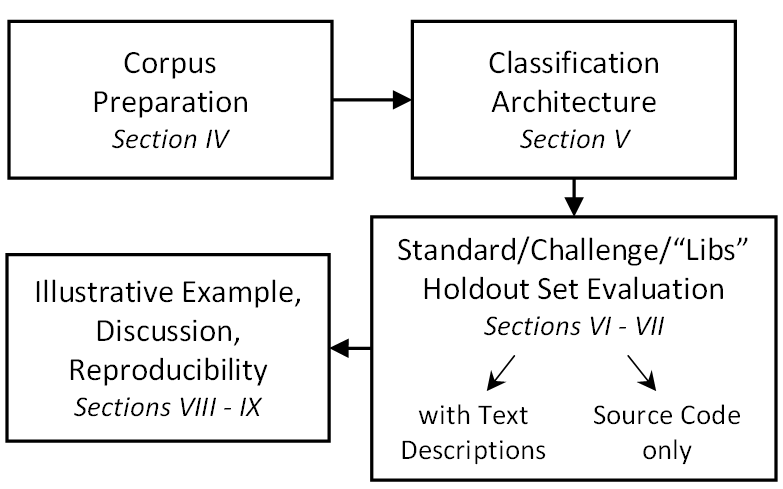}
	\caption{An overview of this paper.}
	\label{fig:overview}
	\vspace{-0.2cm}
\end{figure}

This paper has four major components as shown in the overview in Figure~\ref{fig:overview}.  First, in Section~\ref{sec:dataprep}, we prepare a corpus comprised of C/C++ projects from the Debian \texttt{packages} repository, totaling over 1.5 million files and 6.6 million functions.  The repository contains a special category labeled ``libs'', which we remove and annotate manually for a separate evaluation.  Next, in Section~\ref{sec:approach}, we describe our custom classification approach that is based on neural text classification algorithms described in relevant NLP literature.  Third, in Sections~\ref{sec:eval} and ~\ref{sec:evalresults}, we present our evaluation of our classification approach in comparison with an alternate SE-specific software classification approach, as well as recent work from the area of text classification.  Finally, in Sections~\ref{sec:discussion}~and~\ref{sec:reproducibility}, we present an example illustrating the intuition behind our results, and information for reproducing our work.

%% file: background.tex
\section{Background and Related Work}
\label{sec:background}

This section discusses background technologies including neural text classification, plus key related work.

\subsection{Related Work}
\label{sec:related}

The primary area of related work is software categorization, specifically classification of projects into pre-defined categories in a software repository.  One example of closely-related work is by Linares~\emph{et al.}~\cite{linares2014using}.  Basically, the paper uses terms and API calls as attributes to classify software projects into one of 22 categories in a SourceForge repository of about 3,200 projects.  The paper studied several algorithms, but by far the best result was achieved by a linear classifier using the bag-of-words model, after selecting features using expected entropy loss.  Limiting the attributes to API calls allows binaries to be categorized without code, but was found to impose a cost on performance versus an older approach that treats source code as text and uses text classification algorithms~\cite{Ugurel:2002:WCA:775047.775141}.  However, we caution that the results from those papers are not comparable to our work in this paper, because the feature selection phase using expected entropy loss was performed on the entire dataset, including the test set.  Even though that may be useful for answering interesting research questions about software, it is not a real-world categorization scenario because one would not have the categories of the test set on which to perform feature selection, prior to making the predictions.

Several approaches have been explored in this vein, such as by Sharma~\emph{et al.}~\cite{Sharma:2017:CGR:3084226.3084287}, Kawaguchi~\emph{et al.}~\cite{kawaguchi2006mudablue}, Sandhu~\emph{et al.}~\cite{sandhu2007approaches}, Kim~\emph{et al.}~\cite{kim2018software}, and Yang~\emph{et al.}~\cite{yang2012lacta}.  Wang~\emph{et al.}~\cite{wang2013mining} explore hierarchical categorization, in which whole software projects are placed into a hierarchy e.g. \texttt{players} or \texttt{editors} within the \texttt{video} category.  The overall message from these papers is that a linear classifier using a bag-of-words model of software is sufficient to obtain reasonable performance, when high-level text descriptions are available.  In fact, this conclusion is supported by relevant work in NLP, finding that BoW+LR (bag of words + linear regression) is a strong overall baseline~\cite{wang2012baselines}.

After categorization into pre-defined categories, unsupervised approaches are the next-most similar area of related work.  Unsupervised approaches are quite popular: Chen~\emph{et al.}~\cite{chen2016survey} present a thorough survey of using topic models to organize software in software repositories such as automated Github tagging~\cite{cai2016greta, tian2009using}, and this work is complementary to work to calculate software similarity in terms of features the software implements~\cite{McMillan:2012:DSS:2337223.2337267, ye2016word, zhang2017detecting, gu2016deep} (see a comparison of various similarity metrics by Ragkhitwetsagul~\emph{et al.}~\cite{ragkhitwetsagul2017comparison}).

Broadly speaking, this paper is related to the field of automated program comprehension for software maintenance.  Source code summarization, for example, seeks to generate and update documentation by automatic understanding of software components~\cite{fowkes2017autofolding, mcburney2016automatic, rodeghero2015eye, loyola2017neural}.  Topics such as code summarization, deobfuscation of source code~\cite{vasilescu2017recovering}, bug localization~\cite{lam2017bug, zhou2012should, liu2005sober}, feature location~\cite{dit2013feature}, traceability link recovery~\cite{mills2017automating} are all, at a high level, instances of the concept assignment problem~\cite{biggerstaff1994program} in automatic program comprehension.  A distinguishing factor of this paper is that we target categorization of functions in source code to provide a more fine-grained view software, but without connecting code functionality specific to only a few projects.

\subsection{Neural Text Classification}
\label{sec:ntc}

Neural text classification techniques are usually thought of as in contrast to classical techniques.  In a classical text classification technique, a feature engineering step is followed by feature selection and then training of a machine learning algorithm~\cite{sebastiani2002machine, lai2015recurrent}.  The feature engineering and selection phases are characterized by human design effort, such as choosing a Bag of Words (BoW) model in which each word is a feature, and filtering out words based on entropy loss~\cite{linares2014using}.  There are many choices for ML algorithm, with, for example, Linear Regression (LR) being quite popular~\cite{sebastiani2002machine, Wang:2012:BBS:2390665.2390688}.

Neural classification algorithms have in recent years been proposed as alternatives.  A key advantage to neural algorithms is that they avoid much of the feature engineering/selection work that is dependent on educated guesswork by the designer.  Instead, a network learns the feature representation called a word embedding.  The word embedding, in a nutshell, is a vector space in which each word is a vector of real numbers.  In a neural structure, the embedding can be created either via a supervised training process, or via an unsupervised, pre-trained word embedding based on word co-occurrences~\cite{pennington2014glove}.  Supervised embeddings have the advantage of customization to a particular dataset, but then are limited to that dataset.  Unsupervised embeddings have the advantage to learn from a much larger and more diverse dataset (we use unsupervised embeddings later in this paper for this reason).  

The literature describes a huge variety of neural architectures for text classification~\cite{john2017survey, allahyari2017brief, raju2017comparative}.  In general though, most architectures include convolution and/or recurrent layers combined in different ways for different tasks.  At a high level, a convolution layer (and associated pooling strategy) is used to select which words are the most important, while a recurrent layer is intended to capture the semantics of a word by capturing the context in which a word appears (i.e., the other words to the right or left of a word in a sentence).

In adapting neural architectures to software engineering data, it is worth noting that a large number of the text classification algorithms in existence have been designed for and evaluated on datasets of sentences that have been manually placed into a small set of relatively distinct categories.  For example, the problem of sentiment analysis is quite common: sentences from e.g. movie reviews are labeled as either positive or negative~\cite{socher2013recursive}.  Or, newsgroup categories such as rec.motorcylces versus comp.sys.ibm.pc.hardware~\cite{Lang95}.  It is not surprising that performance levels are reported to be quite high, in the 96\%+ range~\cite{lai2015recurrent}, since a convolution layer will quickly pick out words like ``excellent'' for the positive category, and a recurrent layer can detect important word orders such as ``not bad'' versus ``very bad'', among other semantic factors.

The software categorization problem is different, because 1) the artifacts tend to be much longer than a sentence (in terms of number of words), 2) the vocabulary tends to be different among different projects (due to specialized identifier names), and 3) the categories tend to overlap and are often not polar opposites like negative and positive.  Software categories such as \texttt{database} and \texttt{web} are technically distinct, but are likely to be closely bound together in practice, and are definitely not mutually exclusive.  For these reasons and others, there is significant debate as to whether neural architectures are appropriate for software engineering data at all~\cite{hellendoorn2017deep}, despite their popularity~\cite{allamanis2017survey}.  In Section~\ref{sec:approach}, we will present our argument that a neural approach is an appropriate tool for software classification, and our SE-specific adaptations.

%% file: dataprep.tex
\section{Corpus Preparation}
\label{sec:dataprep}

We prepared a corpus from a snapshot of the Debian \texttt{packages} repository taken September 19, 2017.  The snapshot includes 24,598 software projects.  From these, we filtered projects that did not have any C or C++ source files (some projects contain source files of multiple languages, but we used only the C/C++ components; we chose C/C++ because it was the largest language family in the repository), leaving 9804 projects.  Using a suite of C/C++ analysis scripts provided by the authors of~\cite{mcmillan2013portfolio}, we extracted all files, functions, and function calls. 
We took the category for the project from the Debian package configuration files.  The configuration files allow package maintainers to place a project in more than one category, but in practice we found this to be quite rare, with only a tiny percent of projects having multiple categories.  Since multiple categories per project was unusual, we decided to treat the classification problem as single-class.  For the few projects with multiple categories, we chose only the first category listed in the configuration file.  There were 75 categories in the dataset, though most are very small.

We then created five datasets:

\textbf{Full Holdout/Test Corpus}  We randomly selected 320 projects from the 16 largest categories (twenty from each category) to hold out from all other datasets.  We use these projects exclusively as a testing set in our experiments.  The holdout set is neither used to create the word embedding nor to train the neural architecture.  We kept the holdout to around 5\% of the challenge corpus, to limit removal of training data.

\textbf{Standard Training Corpus}  We intend the standard corpus to reflect the common practice in text classification.  Very widespread practice in the text classification literature is to pare down large, complex datasets into smaller, simpler ones by keeping only the top few categories.  For example, work co-authored by Yann LeCun at NIPS'15~\cite{zhang2015character}, already with over 450 citations, selects the four largest out of 28 categories in an English news corpus to ensure at least 31,900 samples per category, and five out of over 100 categories in Chinese to ensure 112,000 samples per category.  To follow this practice, we create a ``standard corpus'' from the largest six categories (excluding \texttt{libs}, which we treat separately).  Admittedly the choice of six is somewhat arbitrary, but the top six categories cover about 33\% of the projects (excluding libs), making a significant subset.  The six categories are:  \texttt{admin}, \texttt{games}, \texttt{net}, \texttt{science}, \texttt{sound}, and \texttt{utils}.  There are 1,443,408 functions over 3,119 projects in the standard corpus.  To create a balanced set, following an undersampling process similar to~\cite{zhang2015character}, we randomly selected 150k functions (roughly the size of the smallest category) from each category, for a total of 900k functions.

\textbf{Challenge Training Corpus}  In reality, software categorization takes place over a larger and more diverse set of categories than are in the standard corpus.  There is also much more noise, as more categories overlap (a project could be both \texttt{net} and \texttt{web}, for example, even if they only labeled as one category in the dataset), and the categories vary much more in size.  The result is a much more difficult classification problem.  To reflect this realistic situation, we select the top 16 categories (top 17 minus \texttt{libs}), which covers about 50\% of the projects, to form a ``challenge corpus.''  The remaining 58 categories that we do not use cover around 20\% of the dataset.  In our view, categorization of the 58 smallest categories is too far from the capabilities of current technology at this time.  The challenge corpus contains 2,951,529 functions over 5,958 projects.  From these we selected 60k functions (roughly the size of the smallest category) from each category for a total of 960k functions.

\textbf{Libs Holdout/Test Corpus}  We randomly selected 100 projects from the \texttt{libs} category to serve as a holdout/test set for experiments on that category.  We kept the ``libs holdout'' set relatively small (around 5\% of the training set) to facilitate manual annotation and inspection.


\textbf{Libs Training Corpus}  The \texttt{libs} category in the Debian packages repository is a special category because it contains projects that are very likely to also exist in another category, even though in practice the other category is not listed in the repository.  For example, a project may not only be a library, but a library for decoding mp3 files, so it can be considered both in \texttt{libs} and \texttt{sound}.  As a result, \texttt{libs} is a diverse and large category that includes over 30\% of projects.  For this reason, we separate \texttt{libs} from the rest of the repository, annotate it manually with the categories from the rest of the project, and study how well our model classifies the libraries as e.g. a sound library.  See Section~\ref{sec:evalmethodology} for details.


%% file: approach.tex
\section{Our Approach}
\label{sec:approach}

This section describes our approach, including how we represent functions, the classification model we implement, our rationale, and implementation details.  In a nutshell, our approach is to 1) represent each function as a vector of integers and assign a label, 2) train a word embedding for code, 3) train a neural classification algorithm using the function representations and labels from the first step, and 4) to classify a whole project, we classify each of the functions from that project and then apply a voting mechanism to predict a label for the project.

\subsection{Function Representation}
\label{sec:funrep}

We represent each function as a vector of integers.  To obtain this vector, we first create a text string of the function in the form: \texttt{projectName functionName functionContents}.  We convert the project and function names to lower case but do not split or otherwise preprocess them.  For the function contents, we take the code of the function including comments, convert non-word symbols e.g. brackets and commas to spaces, and convert to lower case.  Then we split the entire string on whitespace to obtain a vector of text tokens (which may or may not be English words).  Then, we create an index where each token has a unique integer (to avoid collisions, we do not use the hashing trick).  We use the index to create a vector of integers from each vector of tokens.

Our rationale for including the project name is described in the next section, as it is connected to our use of a recurrent layer.  Note however that in our experiments, we divide the training and testing sets by project, \emph{not} by function, to avoid a situation where the model simply learns which project name belongs in which category (i.e. functions from projects in the training set cannot be in the testing or validation sets).

Our rationale for including the function name is twofold.  First, the function name often includes valuable semantic information~\cite{Hill:2009:ACS:1555001.1555039}.  Second, many projects in our dataset are dependencies of other projects, and in this situation function names from one project will occur in the function contents of functions from other projects (since a project will typically call functions in the projects on which it depends; preserving these call relationships is also why we do not do aggressive identifier splitting during preprocessing).

The function representation above involves information only from the source code.  We refer to it as the \textbf{\texttt{code-only}} or \texttt{co} representation.  The Debian packages repository contains a short (usually about 3 sentences) high level project description for most projects.  Where that description is available, we append it to the code only representation of each function to create a second, \textbf{\texttt{code-description}} or \texttt{cd} representation.  This distinction is important because we use both in the training set: we train our model using both the \texttt{cd} examples and the \texttt{co} examples, to improve the versatility of the model.  The model will learn to classify both situations where text descriptions are available, and where they are not.

\subsection{Training a Word Embedding}
\label{sec:wordemb}

We use three pre-trained word embeddings in this paper.  We used the GloVe word embedding technique~\cite{pennington2014glove}.  Due to space limitations we defer to the paper by Pennington~\emph{et. al}~\cite{pennington2014glove} for a complete explanation, but essentially what the technique does is create a vector space in which words are modeled as vectors, and words that cooccur within a window size (typical is within 15 words) are considered more similar and thus made to be closer in the vector space.  We used a 100 dimension vector space for all three embeddings in this paper.

\textbf{Wikipedia Embedding}  The first embedding we use is the 100d, 6B token word vectors provided by the authors of the original GloVe paper.  These vectors were trained using general text data from the entire corpus of sentences on Wikipedia.  It represents a typical embedding for use in classifying text data.

\textbf{Code-Only Embedding}  The second set of word vectors we used is one that we trained ourselves.  Using a window size of 200 tokens measured as left-context of each word, we computed a word embedding using each function in the training set as a ``sentence.''  Our idea was to produce an embedding in which terms from code are nearby, if they cooccur in functions.  A similar idea has been proposed as Python2Vec~\cite{python2vec}, except that we increase the window size to 200 due to the long length of functions compared to sentences.

\textbf{Code-Description Embedding}  The third embedding we used is also one we trained ourselves, and is, to our knowledge, a novel strategy.  Essentially what we do is \emph{prepend} the project description (the same one used to create the \texttt{cd} function representation) to every function before computing the word vectors.  The description appears before the function in each sequence, so that the words from the description will occur in the left-context of the terms in the function's code.  The idea is that the terms from code will be near in the vector space to words in the high level description.  Our intent is for this to partially address the vocabulary problem, because low-level code terms will cluster around high-level descriptive words.

\begin{figure}[!b]
	\vspace{-0.5cm}
	\begin{small}
		\begin{verbatim}
		01  model = Sequential()
		02  model.add(Embedding(vocab_size, embed_dims,
		03      weights=[embed_matrix], 
		04      input_length=seq_len, trainable=False))
		05  model.add(Conv1D(filters, kernel_size,
		06      padding='valid', activation='relu',
		07      strides=strides))
		08  model.add(MaxPooling1D())
		09  model.add(LSTM(lstm_units))
		10  model.add(Dense(hide_u, activation='relu'))
		11  model.add(Dropout(dropout_level))
		12  model.add(Dense(num_categories,
		13      activation='softmax'))
		14  model.compile(
		15      loss='categorical_crossentropy',
		16      optimizer='adam', metrics=['accuracy'])
		\end{verbatim}
	\end{small}
	\vspace{-0.3cm}
	\caption{Keras code implementing our model, included for maximum clarity and reproducibility.  Also see Section~\ref{sec:reproducibility}.}
	\vspace{-0.1cm}
	\label{fig:modelcode}
\end{figure}

\subsection{Neural Classification Model}

For maximum clarity and reproducibility, we describe our neural classification model in the context of the actual Keras code that we wrote to implement our model (Figure~\ref{fig:modelcode}).  The rapid proliferation of a large variety of neural classification algorithms available makes it quite difficult to select a ``single best'' algorithm, so we designed our own model that characterizes the advancements that seemed broadly effective during our literature review (see Section~\ref{sec:ntc}), centering around convolution and recurrent layers.  Our model is similar to the C-LSTM model that was shown to have performance in line with competitive models on several text datasets~\cite{zhou2015c}.


\vspace{0.1cm}
\paragraph{Our Model}
Our model consists of the following:

\emph{Embedding Layer}  First we use an embedding layer in which every token in the entire vocabulary is represented as a real-valued vector of length \texttt{embed\_dims}.  We used one of the pre-trained embeddings described in the previous subsection, depending on the experiment in future sections, though in general we recommend using the code-description embedding.  Note that we do not eliminate tokens that occur rarely, since these tokens may have useful semantic information, and the convolution layer should deemphasize lesser important tokens anyway.  We used a sequence length of 60 as a compromise between maximizing information available to the model and minimizing model size in memory.  88\% of functions were shorter than our sequence length; we truncated longer sequences.  The output of the embedding layer is a representation of a function that is an $X$ by $Y$ matrix where $X$ is the sequence length and $Y$ is the number of embedding dimensions.

\emph{Convolution Layer}  The next layer is a one-dimensional convolution layer.  Our rationale for using this layer is so that the model would learn which tokens and ``phrases'' of tokens are the most important in determining whether a function belongs in a category.  In text classification, convolution may find phrases such as ``quite well filmed'' to indicate a positive movie review.  In software classification a phrase could be e.g. ``if not muted playsound'' (we do not select these phrases, the convolution layer learns them, see Section~\ref{sec:discussion}).

\emph{Max Pooling}  We used a maximum pooling strategy to downsample the output of the convolution layer and focus on only the more-important tokens.

\emph{Long Short Term Memory}  Next we used a recurrent layer to capture the semantics of tokens in terms of the order in which the tokens appear in the function representations.  We used LSTM due to its ability to capture semantics over a relatively long sequence, which is important because important tokens in our function representation may not be near each other in the sequence.  For example, consider a situation where the model ``sees'' a function representation with a project name \texttt{projA} followed later in the sequence by \texttt{volume}, and that function is in the sound category.  If later during training the model sees a function with \texttt{projA} followed by \texttt{audio}, the model will learn not only that volume and audio are tokens associated with the label sound, but that volume and audio are associated with each other because of their co-occurrence with projA.  The output of the LSTM layer is an embedding representation of each function as a real-valued vector of length \texttt{lstm\_units} (note that this embedding is of the function, not the word embedding layer above).

\emph{Dense Hidden}  We use a fully-connected hidden layer after the recurrent layer following the standard procedure of many neural architectures, to provide a layer for learning how the vector representations from LSTM belong to which categories.  As heavily recommended in related literature~\cite{srivastava2014dropout}, we apply dropout as regularization to resist overfitting.

\emph{Dense Output}  Finally, we include an output layer, after which a function is represented as a real-valued vector of length \texttt{num\_categories}.  The index of the highest value in this vector is the predicted category for the function.

\vspace{0.1cm}
\paragraph{Parameter Tuning}  We used grid search to tune the parameters of our model.
We used all the projects in our standard corpus training set, creating a validation set from 5\% of the standard corpus during grid search to avoid using our holdout/test set during tuning, across the eight parameters listed below.  In any given run, we pick the model that performs best on the validation set (Test-for-best-Validation strategy). Ultimately, we settled on the following configuration:f
\begin{table}[!h]
	\centering
	\vspace{-0.2cm}
	\begin{tabular}{llll}
		\texttt{epochs}         & best-of-3		&	\texttt{filters}        & 250 \\
		\texttt{kernel\_size}   & 3     		&   \texttt{strides}        & 1   \\
		\texttt{lstm\_units}    & 100   		&   \texttt{hide\_u}        & 512 \\
		\texttt{dropout\_level} & 0.5 			&	\texttt{optimizer}		& adamax \\
	\end{tabular}
    \vspace{-0.3cm}
\end{table}

\paragraph{Implementation Details}

We implemented our technique using Keras 2.1.2 running TensorFlow 1.4.0.  Hardware included an E5-1650v4 CPU and two Geforce 1070 GPUs.

\subsection{Voting Mechanism}
\label{sec:voting}

We use a plurality voting mechanism to predict a project category from a list of function labels predicted by our classification model.  The concept is very simple: a project gets the label assigned to the most number of its functions (one function one vote).  More complex voting mechanisms, such as those based on function size or importance~\cite{neate2006coderank}, are an area of future work.

%% file: eval.tex
\section{Evaluation}
\label{sec:eval}

This section describes our evaluation of our approach, starting with our research questions, methodology, and justification for the baselines we chose.

\subsection{Research Questions}
\label{sec:rqs}

A major motivation for this paper is the performance penalty associated with using text classification algorithms on source code data.  We designed our proposed approach to reduce this penalty by several adaptations of text classification to source code, and seek to quantify the affect of these adaptations.  Therefore, we ask the following Research Questions (RQs):

\begin{description}
	\item[RQ$_1$] What is the difference in baseline performance from the case when text descriptions are available, to when only source code is available?
	
	\item[RQ$_2$] What is the difference in our proposed approach's performance from the case when text descriptions are available, to when only source code is available?

	\item[RQ$_3$] What is the effect of the word embeddings on the performance of the neural network-based approach?
\end{description}

To be clear, ``our proposed approach'' means the neural-based approach with the code-description embedding (\texttt{nn+cd}).  The baselines are described in the next subsection.

The rationale behind RQ$_1$ is that the baseline text classification algorithms were designed for natural language text, and may not be applicable off-the-shelf to source code due to the vocabulary problem which has long been recognized in software engineering research (see Section~\ref{sec:intro}).  The rationale behind RQ$_2$ is that our proposed approach is also likely to suffer a performance penalty when classifying code-only data, but we seek to quantify this penalty and compare it to the baseline performance.  One way to view RQ$_1$ and RQ$_2$ is as establishing a likely lower and upper bound for performance expectations: in a real-world scenario, if a practitioner has a set of projects to classify, some projects may have text descriptions while other do not.  Finally, the rationale behind RQ$_3$ is that one key difference between our approach and off-the-shelf text classification is how we create a word embedding, and we seek to quantify the effect of that embedding.

\subsection{Baselines}
\label{sec:baselines}

We use two baselines in this paper: 1) a bag of words linear regression (BoW+LR) classifier, and 2) a neural network-based classifier using a word embedding trained on Wikipedia.

We use the BoW+LR classifier for two reasons.  First, a consensus has formed in the text classification literature that it is a strong baseline across a wide variety of datasets~\cite{wang2012baselines}.  Second, the two most closely-related papers on the topic of software categorization (\cite{linares2014using} and \cite{wang2013mining}) both use a linear classifier and a bag of words representation of software.  We used a vocabulary size of the top 1.8k terms, which was the highest that would fit into the 64gb of memory in our workstation, and is slightly more than the 100 per category used by Linares~\emph{et. al}~\cite{linares2014using}.  In the previous work, the project was represented as a matrix of all words in the project.  However, in pilot studies, we found that performance in this configuration was extremely low (10-15\% F1-measure) when EEL was not applied to the testing set (please note caveats in Section~\ref{sec:related}).  So to provide the best apples-to-apples comparison, we ``swapped in'' the BoW+LR algorithm for the NN-based one we propose in our approach: the BoW+LR algorithm received exactly the same training data as the NN approaches, including our function representation and our voting mechanism.  Then we tested using the same testing data.  We felt that this setup would isolate the effects of the algorithm from other factors such as pre-processing and data representation.

We use an NN-based classifier with the Wikipedia-trained word embedding (\texttt{nn+w}, Section~\ref{sec:wordemb}) as a second baseline.  The neural architecture we propose using is unique only in small ways: a huge variety of neural text classifiers exists, and as noted in Section~\ref{sec:approach}, we built an architecture that characterizes architectures that have been shown to have good performance.  What makes our approach novel overall is the adaptations for code data, one of the most important of which is the procedure for creating a code-description word embedding.  We consider the NN-based classifier with the Wikipedia embedding to be a representative state-of-the-art NN-based text classifier, and to isolate the effects of the embedding versus other adaptations such as pre-processing, we conduct experiments in which we change only the embedding.

\subsection{Datasets}
\label{sec:datasets}

We use the three datasets described in Section~\ref{sec:dataprep}: the standard dataset of six categories, the challenge dataset of sixteen categories, and the \texttt{libs} dataset of six categories.

The standard and challenge datasets use the category provided by the Debian packages repository as a goldset of labels for the projects.  We train all models on the same training sets, to isolate the effects of the models from the effects of variations in the dataset.  We reiterate from Section~\ref{sec:dataprep} that the holdout/test sets are used only during our experiments in this and the next section; we never use them during training.

The \texttt{libs} dataset does not include category labels of its own (they are all labeled in the repository as \texttt{libs}), despite the diverse nature of the dataset.  To obtain labels for the libs holdout set, we hired six professional programmers via UpWork to label every project in the holdout set.  We built a web interface that showed the programmers the project name, description, and website link, as well as a list of the top 19 (20 minus \texttt{libs}) categories in the repository.  We allowed the programmers to chose from a list of 19 categories instead of only the top six, to help ensure that they were not ``forced'' to select a category for a project when the project may not actually be a good fit for those top six.  The programmers selected a primary and up to two secondary categories for each project.  The programmers required on average 2.24 hours.

Once all six programmers had completed labeling the projects, we selected a category label for a project by taking the most-common primary label.  We then chose only the projects with a label from the top six categories.  For 72\% of projects, at least three of the six programmers chose the same primary label (there were no ties).  We consider this an encouraging sign for the reliability of our labeling, considering that the programmers were from around the world, chose from 20 categories, and worked independently.  We do not use the secondary label at this time, but metrics such as top-n precision or analogs to pyramid precision~\cite{murray2008summarizing} are an area of future work.

\subsection{Metrics}
\label{sec:metrics}

We use Precision, Recall, and F1 Score to quantify performance.  In general, we use the F1 Score as the key indicator of model quality, since it reflects both precision and recall.  Technically, we calculated all three metrics using the classification\_report functionality in \texttt{sklearn} version 0.19.1, which computes all three metrics for each category, as well as an overall score weighted by the size of each category in the test set.  This overall score is the main one we report in this paper, but full output is available in our online appendix.

\subsection{Methodology}
\label{sec:evalmethodology}

Our methodology to answer RQ$_1$ is to train a BoW+LR model and a \texttt{nn+w} (neural network + Wikipedia embedding) model using the same dataset, and then run two experiments with this model.  In one, we use the \texttt{co} (code-only) test data, and in another we use the \texttt{cd} (code-description) test data.  Note that both test sets have the same functions; the only difference is whether the text description data is available to the model during testing.  We always use the code-description data and the code-only data from the training set during training.

Our methodology to answer RQ$_2$ is similar to our methodology for RQ$_1$, except that we use our proposed approach instead of the baselines.  The idea is to change only one variable at a time; in this case, the classification algorithm.

Finally, our methodology for RQ$_3$ is to use the same experimental setup, except to change only the embedding.  We conduct experiments with the same datasets, changing only between the Wikipedia embedding, the code-only embedding, and the code-description embedding.

\begin{table*}[b!]
	\centering
	\vspace{-0.2cm}
	\caption{\small Overview of the results.  The column \texttt{Train} indicates the training set used: we always use both code-only and code-description function representations from the training set. \texttt{Test} is the test set: \texttt{cd} means code-description, \texttt{co} means code-only.  Recall from Section~\ref{sec:funrep} that \texttt{cd} simulates the situation when we have descriptions, and \texttt{co} simulates when we only have code, but both \texttt{cd} and \texttt{co} test sets have the same functions from the holdout set.  The top four columns are results simulating when no description data is available (\texttt{co}).  The algorithm \texttt{lr 1.8k} is the linear baseline using vocab size of 1.8k, \texttt{nn+w} is the neural architecture with Wikipedia embedding, \texttt{nn+co} is the code-only embedding, and \texttt{nn+cd} is the code-description embedding.  \texttt{P}, \texttt{R}, and \texttt{F} correspond to precision, recall, and f-measure.  \texttt{run} is the run number associated with the downloadable model and full output available in our online appendix; we include it to assist replication of our experiments and use of the models we created (see Section~\ref{sec:reproducibility}).}
	\label{tab:results}
\begin{tabular}{lllllllllllllll}
	&      &                              & \multicolumn{4}{c}{Standard}            & \multicolumn{4}{c}{Challenge}           & \multicolumn{4}{c}{Libs}                \\
	Train & Test & Algo.                        & P  & R  & F  & run                      & P  & R  & F  & run                      & P  & R  & F  & run                      \\ \hline
	cd+co & co   & \multicolumn{1}{l|}{lr 1.8k} & 43 & 33 & 34 & \multicolumn{1}{l|}{11}  & 36 & 21 & 20 & \multicolumn{1}{l|}{21}  & 52 & 37 & 40 & \multicolumn{1}{l|}{31}  \\
	cd+co & co   & \multicolumn{1}{l|}{nn+w}    & 53 & 46 & 46 & \multicolumn{1}{l|}{101} & 17 & 21 & 17 & \multicolumn{1}{l|}{110} & 61 & 49 & 52 & \multicolumn{1}{l|}{201} \\
	cd+co & co   & \multicolumn{1}{l|}{nn+co}   & 58 & 53 & 54 & \multicolumn{1}{l|}{102} & 26 & 26 & 24 & \multicolumn{1}{l|}{120} & 67 & 54 & 57 & \multicolumn{1}{l|}{202} \\
	cd+co & co   & \multicolumn{1}{l|}{nn+cd}   & 61 & 52 & 53 & \multicolumn{1}{l|}{103} & 34 & 28 & 25 & \multicolumn{1}{l|}{130} & 75 & 63 & 66 & \multicolumn{1}{l|}{203} \\ \hline
	cd+co & cd   & \multicolumn{1}{l|}{lr 1.8k} & 68 & 64 & 65 & \multicolumn{1}{l|}{12}  & 51 & 51 & 48 & \multicolumn{1}{l|}{22}  & 61 & 49 & 52 & \multicolumn{1}{l|}{32}  \\
	cd+co & cd   & \multicolumn{1}{l|}{nn+w}    & 80 & 77 & 78 & \multicolumn{1}{l|}{104} & 44 & 42 & 40 & \multicolumn{1}{l|}{140} & 58 & 49 & 48 & \multicolumn{1}{l|}{204} \\
	cd+co & cd   & \multicolumn{1}{l|}{nn+co}   & 84 & 75 & 76 & \multicolumn{1}{l|}{105} & 49 & 47 & 44 & \multicolumn{1}{l|}{150} & 75 & 66 & 69 & \multicolumn{1}{l|}{205} \\
	cd+co & cd   & \multicolumn{1}{l|}{nn+cd}   & 86 & 80 & 81 & \multicolumn{1}{l|}{106} & 53 & 47 & 46 & \multicolumn{1}{l|}{160} & 76 & 71 & 72 & \multicolumn{1}{l|}{206}
\end{tabular}
\end{table*}

\subsection{Threats to Validity}
\label{sec:threats}

Like any paper, this evaluation carries threats to validity.  One threat in this paper is the selection of dataset.  Due to the computational expense of training and testing each word embedding and model configuration, we pre-select a holdout/test and training set rather than conducting an e.g. 10-fold cross-validation.  While this is the typical strategy in text classification (such as in~\cite{zhang2015character} as mentioned in Section~\ref{sec:dataprep}), it does open up a threat to validity in that results may vary if a different holdout set were used.  We attempted to mitigate this threat by conducting experiments over three datasets.

Another threat to validity is the selection of labels for projects.  There is likely to be overlap in the categories (e.g., between \texttt{net} and \texttt{web}), or a project could be miscategorized in the goldset, so it is possible that some projects are categorized correctly by the models, even if that prediction is calculated as a miss.  We tried to mitigate this threat by using both reference labels from Debian packages, as well as a manually-labeled libs dataset that we held out completely from training and testing with the reference labels.


%% file: evalresults.tex
\section{Evaluation Results}
\label{sec:evalresults}

We present our answer to each research question in this section, with the supporting data and rationale.  We offer an interpretation of each result, which we discuss further in the following section.

\subsection{RQ$_1$: Baseline Performance}
\label{sec:r1results}

\begin{figure}[!t]
	\centering
	\includegraphics[width=0.43\textwidth]{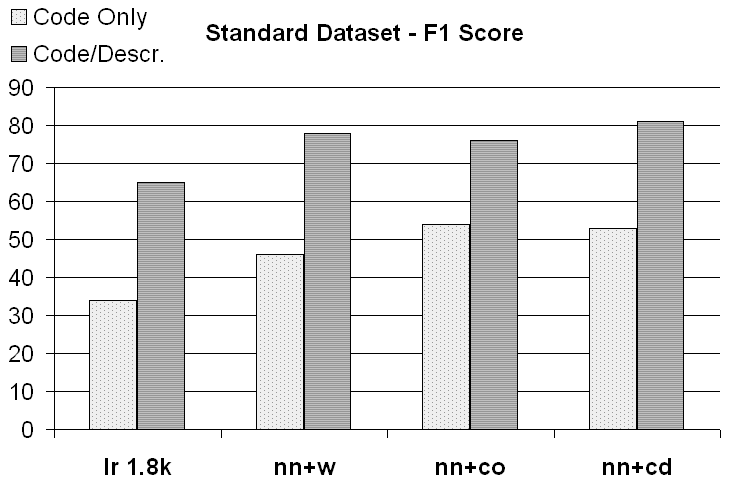}
	\caption{\small F1 Score for code-only and code-description standard datasets.  We observe a significant drop in performance from code-description to code-only in all cases, though this drop was lower for \texttt{nn+cd} than \texttt{nn+w}.  Best overall performer was \texttt{nn+cd}.}
	\label{fig:std_f1}
	\vspace{-0.5cm}
\end{figure}

We observe that both baselines suffer a significant drop in performance, when faced with code-only data versus code-description data, as clearly visible in Figure~\ref{fig:std_f1}.  Consider first lines 1 and 5 in Table~\ref{tab:results}.  These lines correspond to the linear regression for code-only (line 1) and code-description (line 5) data.  In line 5, the F1 score on the standard dataset is 65, and precision/recall is 68/64.  These results are broadly similar to the 67\% F1 score reported by Wang~\emph{et. al}~\cite{wang2013mining} when using text description data (mined from the web) with linear regression to classify Java projects from SourceForge (though the results are not directly comparable due to different categories, datasets, and programming language).  However, when text data is not available, we observe a drop in performance to 34\% F1 score, 43/33 precision/recall -- a nearly 50\% performance penalty as measured by F1 score.

The same pattern is visible on the challenge and libs datasets.  On the challenge data, the F1 score is 48 for \texttt{lr} with descriptions, but only 20 without.  And on the libs data, the F1 score is 52 with descriptions versus 40 without.  Performance is consistently much lower on code-only data.

The baseline neural-based approach with the Wikipedia embedding (\texttt{nn+w} in Table~\ref{tab:results}) exhibits similar behavior.  It achieves 78\% F1 score with descriptions, compared to 46\% without -- over a 40\% performance drop.  Challenge results are even more extreme, from 40 to 17 F1 score.  Interestingly, performance on the libs dataset is relatively stable, even with slightly improved results when description data is removed: from 48 to 52.  One likely explanation is that the library functions could be called by functions in the training set, which would lead to the model learning to recognize those libraries as part of a category.  Still, on the standard and challenge datasets, there is evidence for a large performance gap between code-only and code-description data.

This result may not be all that surprising, in light of the algorithms' origin in the NLP literature.  The more-recent neural-based approaches were designed to exceed baseline LR performance on text data, and we observe that they do with text descriptions of software on the standard dataset.  But that is basically a text classification problem that happens to be using text data from the software domain.  Baseline performance of \texttt{lr} and \texttt{nn+w} tends to be low on both the standard and challenge datasets for code-only test data.  Our interpretation of these results is that classifying source code is a different problem than text classification, highly likely to be due to the vocabulary problem.  In short, we find that an off-the-shelf application of text classification approaches to source code data is not likely to be effective.

\vspace{-0.2cm}
\subsection{RQ$_2$: Proposed Approach Performance}
\label{sec:r2results}

Lines 4 and 8 correspond to the code-only and code-description results for our proposed approach (\texttt{nn+cd}).  We observe a drop from 81 F1 score for \texttt{cd} to 53 for \texttt{co} on the standard dataset.  This is roughly a 35\% drop -- certainly very significant, but the effect is not as pronounced as in the baseline.  In fact, the \texttt{nn+cd} result on code-only data is low-end competitive with the baseline performance on code-description data.  On the standard dataset, F1 score for \texttt{nn+cd} was 53 for code-only, compared to 65 for \texttt{lr} code-description.

The challenge dataset results are more mixed.  While \texttt{nn+cd} does obtain the highest F1 score on both \texttt{co} and \texttt{cd} groups among the neural-based approaches, the \texttt{lr} baseline actually has the highest performance, though by a small margin.  Given the threat to validity we mention in the previous section of avoiding cross-validation due to computational expense, it is difficult to draw a strong conclusion from the challenge set.  Also, we note that the performance penalty for \texttt{nn+cd} is roughly 50\% on the challenge dataset, similar to the baseline.

Remarkably, we observe little difference between \texttt{nn+cd} performance on the libs dataset.  As with baseline performance, we surmise that the reason is that the libraries are used by the end-user programs, which means that the training set will include function names of the libraries in the code-only data; the high-level descriptions may not add much information.

\subsection{RQ$_3$: Effects of Embeddings}
\label{sec:r3results}

We found evidence that the embeddings had an important impact on the performance of the neural-based approach -- specifically, that \texttt{nn+cd} was the best performing approach overall.  Comparing \texttt{nn+cd} to \texttt{nn+w}, we observe an increase of 3 points on F1 score on the \texttt{cd} standard dataset, and an increase of 5 points on the \texttt{co} standard dataset.  The performance increase for \texttt{nn+co} and \texttt{nn+cd} over \texttt{nn+w} is larger on the code-only data, probably because the embeddings include words from the code, while the Wikipedia embedding is less likely to have the same vocabulary as is used in code.  We will demonstrate an example of the benefits conferred by the embeddings in the next section, but for now we observe an increase in performance, which is larger on code-only data.

The difference between the embeddings is clearer on the \texttt{libs} dataset.  F1 score for \texttt{nn+cd} is 72 with text descriptions, compared to 48 for \texttt{nn+w}.  With code-only, \texttt{nn+cd} achieves 66 F1 score versus 52 for \texttt{nn+w}.

But as with RQ$_2$, it is difficult to make strong conclusions from the challenge dataset when comparing the embeddings.  The \texttt{nn+cd} configuration (our proposed approach) does perform slightly better, but the improvement is large only when comparing \texttt{nn+cd} to \texttt{nn+w}.  While an area of future work is verifying the difference on multiple holdout sets (time permitting), our interpretation is that the challenge dataset is near the edge of the capability of the algorithms we discuss.  One possible reason is that all approaches (\texttt{lr} included) are picking up on the ``easy'' clues for each class, while missing more subtle details.  One remedy may simply be to include more training data (perhaps by oversampling instead of undersampling during corpus preparation, Section~\ref{sec:dataprep}), but it seems likely that work will need to be done in designing classification models sensitive to the specifics of software.

Overall, we find that \texttt{nn+cd} is the best performer, at least by a small margin, in all situations except code-only on the standard set, where it is ahead on precision but lags \texttt{nn+co} on recall by one point (a very small difference).

%% file: discussion.tex
\begin{figure*}[t!]
	\centering
	\begin{subfigure}[b]{\textwidth}
		\includegraphics[width=\textwidth]{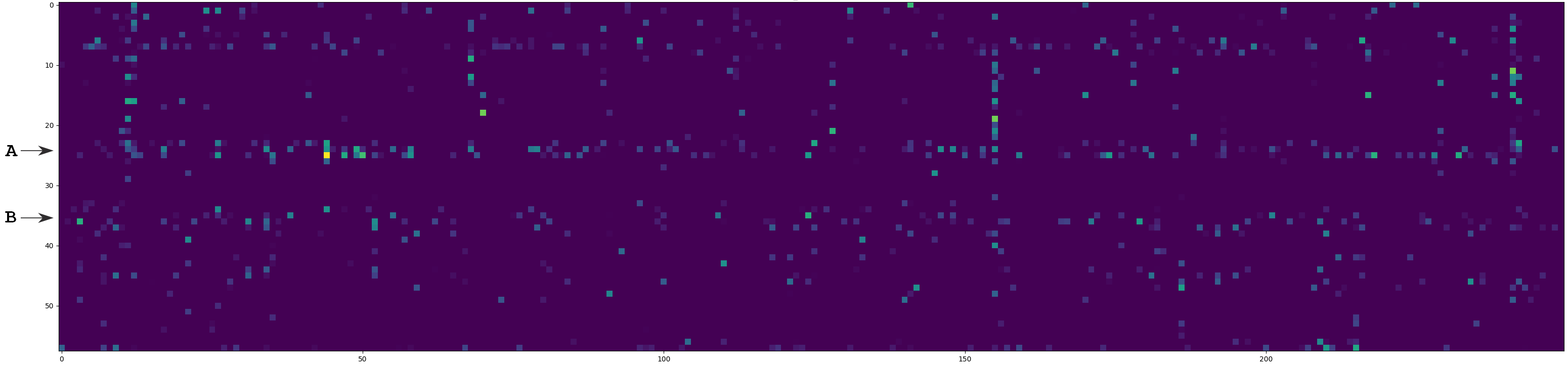}
		\vspace{-0.5cm}
		\caption{NN+W (Wikipedia embedding), predicting \texttt{science} category (incorrect).}
		\label{fig:w_6}
	\end{subfigure}
	
	\vspace{0.3cm}
	
	\begin{subfigure}[b]{\textwidth}
		\includegraphics[width=\textwidth]{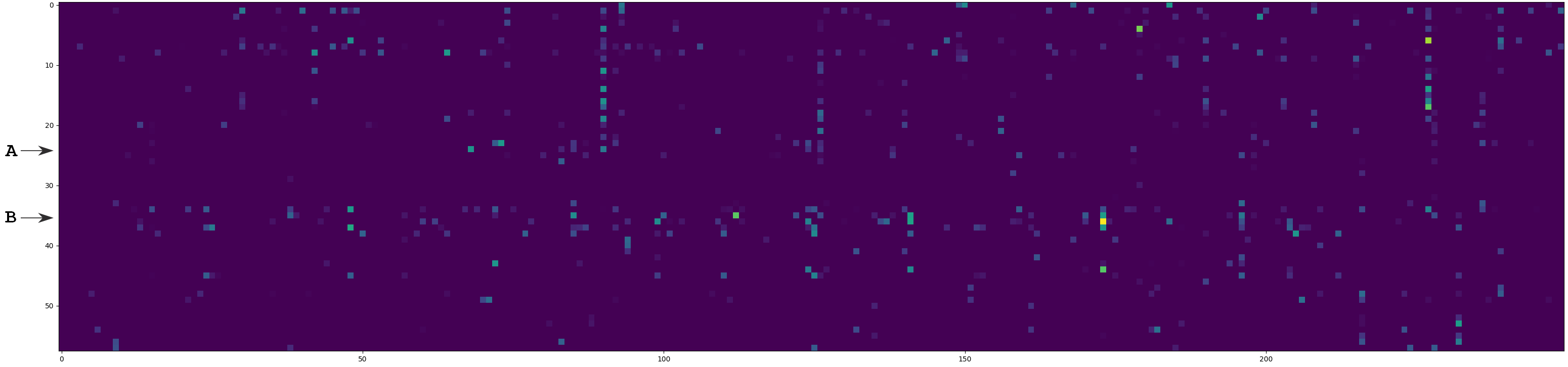}
		\caption{NN+CD (Code-Description embedding), predicting \texttt{utils} category (correct).}
		\vspace{-0.1cm}
		\label{fig:cd_6}
	\end{subfigure}
	
	\caption{\small Heatmaps of the attention of the convolutional layer in the neural-based approaches.  The only difference between (a) and (b) is the embedding.  Both were trained with the same training data and using the same parameters.  We generated the maps using the function in Figure~\ref{fig:starcode}.  The vertical axis represents the tokens (60 sequence length) with the first token at the top.  The horizontal axis represents the 250 filters in the convolution layer.  Arrows denote areas A and B, which are discussed in Section~\ref{sec:discussion} and shown in Figure~\ref{fig:heatmaps}.}
	\label{fig:heatmaps}
	\vspace{-0.4cm}
\end{figure*}

\begin{figure}[!b]
	\vspace{-0.4cm}
	\begin{small}
		\begin{Verbatim}[commandchars=\\\{\},codes={\catcode`$=3\catcode`_=8}]
stardict q q caleditdistance n s m temp itemp
currentelements k n d k m d i n j i cost k
cost d \textcolor{Fuchsia}{\textbf{compute levenshtein distance}} between s
and t this is using quick \textcolor{PineGreen}{\textbf{algorithm descrdelim}}
\textcolor{PineGreen}{\textbf{international dictionary}} stardict is a cross
platform and international dictionary written
in gtk 2 x it has powerful features such as
glob style
		\end{Verbatim}
	\end{small}
	\vspace{-0.1cm}
	\caption{\small Code-Description representation of the function CalEditDistance from StarDict, a project in the \texttt{utils} category.  Purple text corresponds to area A in Figure~\ref{fig:heatmaps}, and green text to area B.  The descrdelim term is a delimiter we added to denote the separate between words from code and words added from the project description.  The code-only representation would only have terms prior to descrdelim.}
	\vspace{-0.1cm}
	\label{fig:starcode}
\end{figure}

\section{Discussion/Conclusion}
\label{sec:discussion}

In this paper, we proposed several adaptations to off-the-shelf neural-based text classification to the domain of software categorization.  To summarize, these adaptations are:

\begin{itemize}
	\item Represent projects as functions, and assign each function the label of the project from which it originated.
	\item Model functions as a sequence of tokens with minimal pre-processing (Section~\ref{sec:funrep}).
	\item Train a word embedding based on a combination of code and high-level descriptions (Section~\ref{sec:wordemb}).
	\item Use convolutional and recurrent neural layers with the parameters we found via a grid search.
	\item Train the neural model with both code-only and code-description examples of the same functions.
	\item Use a voting mechanism to produce project predictions from the model's function predictions (Section~\ref{sec:voting}).
\end{itemize}

There are numerous areas of future work, including customized neural architectures for different software data, improved voting mechanisms, experiments with different pre-processing techniques, and variations on embedding strategy.  Nonetheless, in experiments with both reference data from Debian user-end programs (the standard and challenge datasets), plus manually-annotated programming libraries, we found improvement over off-the-shelf applications of text classification, as well as a baseline from previous software maintenance literature (RQ$_2$).  We observed that performance for all approaches, but especially the baseline approaches, was heavily influenced by the presence of high-level text data: when only code data was available, performance was far lower (RQ$_1$).  We also found that our proposed code-description embedding was an important contributor to the success of our approach (RQ$_3$).

To begin understanding \emph{why} the code-description embedding achieved higher performance, consider the example function in Figure~\ref{fig:starcode}.  This function belongs to \texttt{stardict}, which is a dictionary program categorized in the reference under \texttt{utils}.  The \texttt{nn+cd} approach correctly predicted the category for this function, but the \texttt{nn+w} approach predicted the function to be part of the \texttt{science} category.  Keep in mind that the only difference between these two models is the word embedding: all parameters and training data were the same.

Figure~\ref{fig:heatmaps} shows a heatmap of the activation of the convolutional neural layer for both approaches.  The horizontal axis is 250 wide: one column for each filter (the parameter value 250 for number of filters was selected after grid search tuning).  The vertical axis is 60 tall: one row for each term (recall that we used a sequence length of 60).  Area \texttt{A} in both heat maps corresponds to terms 25, 26, and 27.  The \texttt{nn+w} model activates heavily across these terms, which as seen in Figure~\ref{fig:starcode}, correspond to the phrase ``compute levenshtein distance.''  In particular, the word ``levenshtein'' leads to significant activation, which implies that that word is an important influencer of the network's decision.

``Levenshtein'' is an unusual word in the code corpus, occurring only a handful of times.  However, in the Wikipedia embedding, the nearest words in the vector space are ``hyperfocal'', ``mahalanobis'', and ``comoving.''  The first two do not appear anywhere in the dataset, but ``comoving'' appears in the function \texttt{normalizeData} in the project \texttt{yt}.  The description of \texttt{yt} is ``an integrated science environment for collaboratively asking and answering astrophysical questions'' -- the \texttt{yt} project is classified as \texttt{science}.  What has likely happened is that the network, via the Wikipedia embedding, connected the word ``levenshtein'' to a project in the science category, based on the word's English usage.  In fact, the English usage is not what matters: the \texttt{nn+cd} approach activates more on other words, such as those in area \texttt{B} and beyond (corresponding to the text ``international dictionary''), which are linked via the embedding to terms in the \texttt{utils} category.  There is not space in this paper to show all these terms, but we provide a Tensorboard interactive visualization of both embeddings via our online appendix.  There is also not space to discuss the categorization results at function-level (we presented project-level predictions, after our voting mechanism), but these results are also available in the appendix.

An important area of our future work is to further understand how code and natural language differ, so that we can make better adaptations of NLP work to the SE domain, a rapidly growing research area~\cite{allamanis2017survey}.  We view this paper as significant progress in that direction.
